# Quasi-guiding modes in microfibers on high refractive index substrate

Kaiyang Wang[1], Zhiyuan Gu[1], Wenzhao Sun[1], Jiankai Li[1], Shumin Xiao[2], Qinghai Song[1, 3]

1. *Integrated Nanoscience Lab, Department of Electrical and Information Engineering, Harbin Institute of Technology, Shenzhen, China, 518055*
2. *Integrated Nanoscience Lab, Department of Material Science and Engineering, Harbin Institute of Technology, Shenzhen, China, 518055*
3. *State Key Laboratory on Tunable laser Technology, Harbin Institute of Technology, Harbin, China, 158001*

## Abstract

Light confinement and amplification in micro- & nano-fiber have been intensively studied and a number of applications have been developed. However, the typical micro- & anno-fibers are usually free-standing or positioned on a substrate with lower refractive index to ensure the light confinement and guiding mode. Here we numerically and experimentally demonstrate the possibility of confining light within a microfiber on a high refractive index substrate. In contrast to the strong leaky to the substrate, we found that the radiation loss was dependent on the radius of microfiber and the refractive index contrast. Consequently, quasi-guiding modes could be formed and the light could propagate and be amplified in such systems. By fabricating tapered silica fiber and dye-doped polymer fiber and placing them on sapphire substrates, the light propagation, amplification, and laser behaviors have been experimentally studied to verify the quasi-guiding modes in microfer with higher index substrate. We believe that our research will be essential for the applications of micro- and nano-fibers.

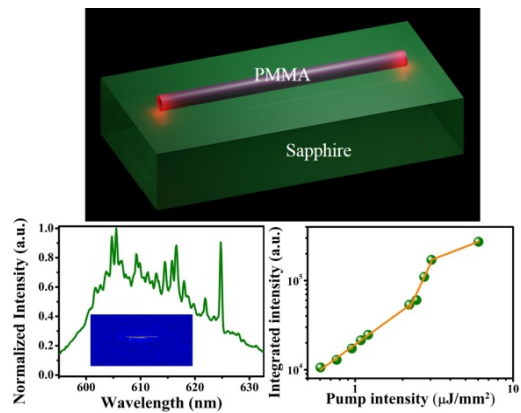

# Introduction

With the rapid developments in nanoscience and nanotechnology, there is a strong demand for interconnection between on-chip micro- & nano- circuits and conventional optical elements. Compared with the conventional devices, microfiber (or nanofiber) has extremely long distance in one dimension and well defined micro-structures (or nano-structures) in other two dimensions [1]. Combined with their intrinsic advantages in low cost, easy fabrication, mechanical flexibility, and compatible integration with typical fiber systems, microfibers thus have been intensively studied as an effective bridge between typical optical elements (such as free space light source) and the micro- & nano- photonic circuits. In past few years, the applications of microfiber (nanofiber) have been further extended to optical couplers [2, 3], filters [4, 5], sensors [6, 7], nano-laser source [8-10] et al.. Typically, the microfiber (nanofiber) is free-standing in air or low refractive index solutions. Consequently, the diameter of microfiber can be significantly reduced to as small as a few nanometers and the interaction between fiber and environment has been dramatically improved. However, the free-standing micro- & nano- fibers have fatal disadvantages. They are extremely sensitive to the environmental vibrations, making the devices to be very unstable [11]. For on-chip devices, the side to side coupling is much easier and more efficient than the end-fire injection [12]. But this kind of coupling requires the light propagate a short distance when the fiber is attached onto the substrate.

Several techniques have been developed to improve the stability of fiber coupling system and the light propagation on substrate [13-15]. To ensure the light confinement within microfiber, the refractive index contrast between fiber and substrate has been considered to a key factor. Experimentally, micro- & nano- sized silica fibers have been fixed on $MgF_2$ substrate [13] or aerogel [14] which have very low refractive indices. For some applications on glass substrate, the refractive index of fiber can also been increased to 2.08 by replacing silica with tellurite [16] or doping. The light can be well confined within micro- & nano- fibers on these substrates. However, the materials of fibers and substrates are strongly limited. For example, in practical applications, the substrates usually have higher refractive indices (e.g. Si, n ~ 3.5), which makes extremely difficult to fabricate microfiber with even higher refractive index. Very recently, slot waveguide and quasi-waveguide have also been proposed to confine light by applying low refractive index materials [17, 18]. But the first one is limited in nanostructure, whereas the latter one is still confining light within the waveguides with higher effective refractive indices. And both of them cannot be directly compatible with conventional fiber system Therefore, it is very interesting and important to explore the possibility of confining and guiding light in low refractive index microfibers on a higher index substrate.

## Results and Discussions

## Theoretical analysis and numerical calculations

In a simplest three-layer slab waveguide, the modes are the solutions of Maxwell's wave equation

$$\nabla^2 E(r,t) = \frac{\left[\frac{n^2(r)}{c^2}\right]\partial^2 E(r,t)}{\partial t^2}, \qquad (1)$$

where E is the electric field vector, r is the radius vector, n(r) is the index of refraction, and c is the speed of light in vacuum. For monochromatic plane waves, the modes can be calculated by taking the boundary conditions. Eq. (1) simply showed that modes couldn't be formed in the region with lower refractive index. This information can be confirmed by a numerical calculation. Two types of numerical methods that are based on finite difference time domain (FDTD) method and finite element method (FEM) have been applied and consistent results have been obtained. One result is shown in Fig. 1(a), where a 1500 nm silica lab waveguide is placed on a sapphire substrate. The refractive indices of sapphire and silica are 1.7676 and 1.473, respectively. While the light can be weakly confined within the waveguide, the effective refractive index is around 1.4609 – 0.0015i for mode at 600nm. The large imaginary part of effective refractive index ($Im(n_{eff})$) corresponds to larger radiation loss. As shown in Fig. 1(a), the numerical results showed that more than 90% energy was distributed in the substrate, consistent with theoretical model well.

Once the cross-section geometry was changed from slab to circle, the light propagation behaviors were totally different. One result is plotted in Fig. 1(b). The refractive indices are the same as Fig. 1(a). And the diameter of circle is also 1500 nm. We can see that the mode is mainly confined within the circular region even though its refractive index is much lower than the substrate. The effective refractive index is changed to 1.4453 – 0.0009i for mode at 600nm. In conventional waveguides, smaller effective refractive index ($n_{eff}$) means weaker light confinements. Here the loss of circular waveguide is actually lower than the slab with larger $n_{eff}$. The percentage of energy in substrate is only as small as 31.9%, much smaller than the leakage of slab waveguide. The reduction of light leakage is mainly caused by smaller contact area between waveguide and substrate, where the total internal reflection is broken. Considering the relative high confinement of light in microfiber, we call such kind of propagation modes as quasi-guiding modes.

To fully understand the losses of quasi-guiding modes, we have calculated the light propagation with full three-dimensional calculations with FDTD method. The simulated structure is depicted in Fig. 2(a), where the diameter of fiber and refractive indices are the same as Fig. 1(b). The corresponding field distribution in side view is shown in Fig. 2(b), where several types of radiation losses can be identified. The first loss is the radiation loss from the mode conversion at position A. When the fiber is positioned in air and on sapphire substrate, their propagation modes are quite different even though both of them are close to fundamental modes. According to the overlap integration, the mode conversion efficiency is only is very low. [19] The second loss is the propagation loss due to the breaking of total internal reflection. By calculating the transmission between point B and C in Fig. 3(b), the propagation loss is around 0.04 dB/μm. While the propagation loss is much higher than conventional micro- & nano-fiber, it is enough to support a short distance of side-side on-chip direct coupling. Moreover, the propagation loss can be further optimized, as shown below.

Since substrate induced leaky energy originates from the refractive leakage, we then focused our attention on the influence of radius of microfiber and the refractive index of substrate. Here the transmitted light wavelength and refractive index of fiber was fixed. As shown in Fig. 1, the leakages of microfiber are strongly dependent on their imaginary parts of effective refractive indices. We thus studied the imaginary part of refractive index as a function of radius and refractive index of substrate. All the results are shown in Fig. 3. As nanofiber radius increases, the imaginary part of effective refractive index increases simultaneously. When the nanofiber radius is larger than 4 μm, the Im($n_{eff}$) approaches zero quickly and the energy inside the substrate is less than 1%. The smaller Im($n_{eff}$) can be easily understood. The modes in Fig. 3(a) are all fundamental modes, whose incident angles increase with the increase of radius. Following the Fresnel equation, the increased incident angle gives higher reflectance. Although the total internal reflection at the fiber-substrate interface is broken, the reflectance at large incident angle is also very close to 100%. In this sense, the total loss is decreased at larger radius. A direct comparison of leakage at different r is plotted in the insets of Fig. 3(a). The leakage at smaller r is quite dramatic and becomes negligible when r is increased to 5 μm.

Figure 3(b) shows the dependence of Im($n_{eff}$) on the refractive index of substrate. In contrast to the intuitive understanding that more energy will leak into substrate with higher refractive index, our calculation shows that Im($n_{eff}$) decrease first and finally approaches to 0 at large refractive index. This non-monotonous behavior is caused by the balance of reflection at the fiber-substrate interface and breaking of total internal reflection. Following above results, we thus know that light can be guided within microfiber even though it was placed onto a higher index substrate. More importantly, the propagation efficiency can be improved when the substrate has much higher refractive index. This can also be directly observed from the cross-sectional field distributions. As shown in the insets of Fig. 3(b), less field leaks from microfiber at larger refractive index of substrate. This thus ensure that microfiber can be positioned onto many types of substrates (see GaN, Sapphire for UV to visible light range, and GaAs, Si for infrared range) to improve the stability and extend the applications in bridging the nanophotonics and traditional optics.

## Experimental Results

Based on above descriptions, we then experimentally fabricated the microfibers to test the light propagation on high refractive index substrate. We first fabricated the silica nanofiber with typical flame-assisted fiber drawing technique and placed it onto a sapphire substrate [1]. Following the calculation in Fig. 2, the propagation loss is acceptable. The microscope image of fabricated microfiber is shown in Fig. 4(a). The diameter of microfiber is around 2.2 μm and it is quite uniform within a long distance. As the microfiber was drawn from typical single-mode fiber, one end of fiber is single-mode fiber. Thus a He-Ne laser can be coupled to fiber through a 10x objective lens. As the propagation loss in single fiber is negligibly small, the light can propagate to the microfiber efficiently. Figure 4(b) shows the corresponding of light propagation within the microfiber. Similar to the simulations in Fig. 2(b), several different leakages can be observed. Due to the non-perfect mode conversion, one leakage is the loss at the position where microfiber reached the substrate. The second one is propagation loss caused by the breaking of total internal reflection. These two leakages can be observed from the light within sapphire substrate. In Fig. 4(b), several bright spots are caused by the scattering at the defects of microfiber. From these spots, it is easy to read out that light propagates well within microfiber even though it is placed onto a higher index substrate. This is consistent with the numerical results.

Based on the observation of light propagation, we then tested the possible light amplification or laser behaviors in such systems. This information is also important for developing efficient coupler or active integrated elements. The microfiber was fabricated by direct drawing from Rhodamine B (RhB) doped Polymethyl methacrylate (PMMA) solution [7]. The PMMA solution was synthesized by solving 1g PMMA and 0.01g RhB into chloroform solvent. The diameter of as-drawn microfiber was controlled by the drawing speed and the concentration of PMMA solution. The fabricated PMMA microfiber was placed onto a sapphire substrate and pumped with a frequency doubled Nd:YAG laser (pulse width 6ns). The pumping laser was focused by a cylindrical lens ($f$ = 25cm) and formed a laser stripe along the axis of microfiber [20]. The width of laser stripe was controlled by a slit. The outputs from microfiber was collected by a lens and coupled to a CCD based spectrometer.

Figure 5(b) shows an example of emission spectrum from microfiber. The pump density is 37 $\mu J/mm^2$. We can see that the linewidth is only around 10nm, which is much smaller than the typical photoluminescence of RhB dye. Figure 5(c) shows the outputs as a function of pump power. With the increase of pump density, we can clearly see a transition from a broad photoluminescence peak to a narrow peak when pump density is larger than 5.85 $\mu J/mm^2$. The corresponding integration of output has been summarized in Fig. 5(d) in logscale. A clear threshold behavior can be observed from the "S" shape [21]. The threshold is around 5 $\mu J/mm^2$. The slope increases from ~ 1 to 2.35 when the pumping power passes the threshold and finally changes back to ~ 1. This "S" shape demonstrates the existence of amplified stimulated emission (ASE) well.

The threshold clearly demonstrates the amplification inside the microfiber. As only the propagating waves along the fiber axis can be amplified, Fig. 5 also gives an additional proof for the existence of quasi-guiding modes when the microfiber is placed onto a higher index substrate. The inset in Fig. 5(b) is the image of nanowire under optically pumping. Consistent with the amplification, we can directly see that most of light is confined along the nanowire, where the intensity is much higher than the others. In the inset of Fig. 5(b), slight emissions from the substrate can also be observed at the edge of substrate (see the dashed line). This kind of emission is mainly caused by the leakage from microfiber, consistent with the numerical calculations in Fig. 2 well. Moreover, as the leakages are not confined by the microfiber any more, large divergent beam instead of directional output at the boundary of substrate was recorded in the images.

In Fig. 5(b), a few spikes can be observed at the peak of ASE [22]. As the linewidth of spikes are smaller than 1 nm, they might relate to lasing actions. We have further increased the pump power to confirm the lasing behaviors. However, no discrete laser peaks have been clearly separated from ASE even though the pump power was close to the damage threshold. We then fabricated another nanowire with the same fabrication process and measured its corresponding outputs. All the results are summarized in Fig. 6. From the microscope image (Fig. 6(a)), we can see that the diameter is about 5 μm, which is much larger than the one in Fig. 5. Following the numerical calculations in Fig. 3(a), the larger diameter relates to better light confinement and low loss. In this sense, the new microfiber might be able to support lasing actions that hadn't been obtained in Fig. 5.

Figure 6(b) shows the emission spectrum the microfiber. Even the pump density is only as small as 6.04 μJ/mm$^2$ (much smaller than that in Fig. 5(b)), randomly distributed spikes can be observed. The number and positions of peaks randomly changes with the decreasing of pumping stripe length, demonstrating random laser behaviors well [23-25]. The lines in Fig. 6(c) are the laser spectra under different pumping density. With the increase of pumping power, the transitions from spontaneous emission to ASE and finally to lasers can be observed. The threshold behaviors can also be observed from the "S" in the logsclae L-I curve in Fig. 6(d). The corresponding laser threshold is around 2.2 μJ/mm$^2$.

The lower threshold indicates that stronger amplification has been achieved in the larger microfiber. Following the numerical analysis, the stronger gain here should be formed by the smaller leakage at larger radius (r). An indirect proof can also be observed in the image of microfiber laser (see the inset in Fig. 6(b)). Similar to the inset in Fig. 5(b), we can see that most of light is confined along the microfiber well. Here, however, the leakages along the substrate are negligibly small and hardly observed at the edge of substrate (marked with dashed lines). One may argue that threshold behaviors might also be caused by different fiber length. To exclude this possibility, we have also checked the slope above threshold. Here the slope is around 3.78, which is much higher than that in Fig. 5(d) and confirms the stronger amplification well.

## Conclusion

We have numerically and experimentally studied the possibility of light propagation in a microfiber that is placed on a higher index substrate. In contrast to the intuitive picture, we found that the fundamental modes could be considered as quasi-guiding modes. The propagating loss of quasi-modes can be minimized by changing the size of microfiber and the refractive index of substrate. And the loss of silica microfiber on silicon is negligibly small. By utilizing the sapphire as substrate, we have experimentally verified the light propagation and amplification in silica and PMMA microfibers. In a larger PMMA microfiber, the lasing actions have also been obtained. Our finding is a new propagating status and can be significantly help the interconnection between conventional fiber optics and micro- or nano-photonics.

## Acknowledgement

This work is supported by NSFC11204055, NSFC61222507, NSFC11374078, NCET-11-0809, Shenzhen Peacock plan under the Nos. KQCX2012080709143322 and KQCX20130627094615410, and Shenzhen Fundamental research projects under the Nos. JCYJ20130329155148184, JCYJ20140417172417110, JCYJ20140417172417096.

**Figure caption:**

Fig. 1: **Quasi-guiding modes on higher index substrate.** (a) Slab silica waveguide on sapphire. (b) Silica Microfiber on sapphire. Here the waveguide is 600 nm. The height and diameter of slab and circle are both 1500 nm.

Fig.2: **Three-dimensional simulation of quasi-guiding mode.** (a) Schematic picture of microfiber on a sapphire substrate. (b) The side view of propagation inside microfiber.

Fig.3: **The dependence of Im($n_{eff}$) on radius of microfiber and refractive index of substrate.** (a) The refractive indices of silica and sapphire are the same as Fig. 1. The insets are the cross-sectional field distributions at different radius (r). (b) The radius and refractive index of silica microfiber are r = 0.5 and n = 1.473, respectively.

Fig.4: **Light propagation in silica microfiber on sapphire substrate.** (a) The microscope image of microfiber on sapphire substrate. The inset shows the diameter is around 2.2 μm. (b) The propagation of He-Ne laser in microfiber.

Fig.5: **Light amplification in PMMA microfiber on sapphire substrate.** (a) The microscope image of dye-doped PMMA microfiber. The diameter of fiber is around 1.7 μm. (b) The emission spectrum of PMMA microfiber. Here the pump power is 37 μJ/mm$^2$. The inset shows the image taken with a conventional camera. (c) The emission spectra from microfiber under different pumping conditions. (d) The logscale input-output curves.

Fig.6: **Random lasers in PMMA microfiber on sapphire substrate.** (a) The microscope image of dye-doped PMMA microfiber. The diameter of fiber is increased 5 μm. (b) The random laser spectrum from PMMA microfiber. Here the pump power is 6.04 μJ/mm$^2$. The inset shows the image taken with a conventional camera. (c) The laser spectra from microfiber under different pumping conditions. (d) The threshold behaviors of random lasers.

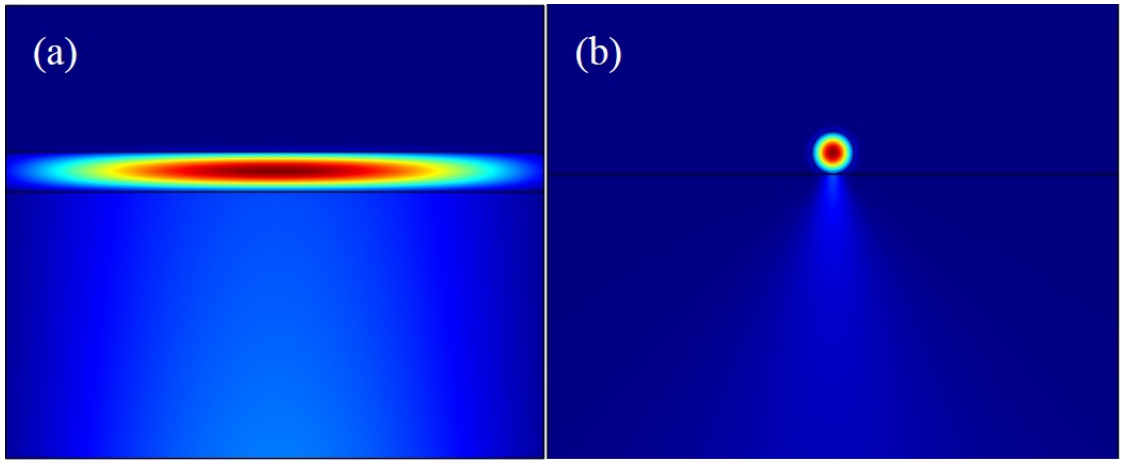

Fig.1

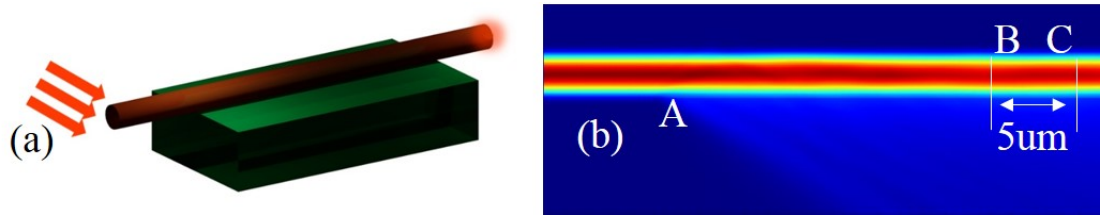

Fig.2

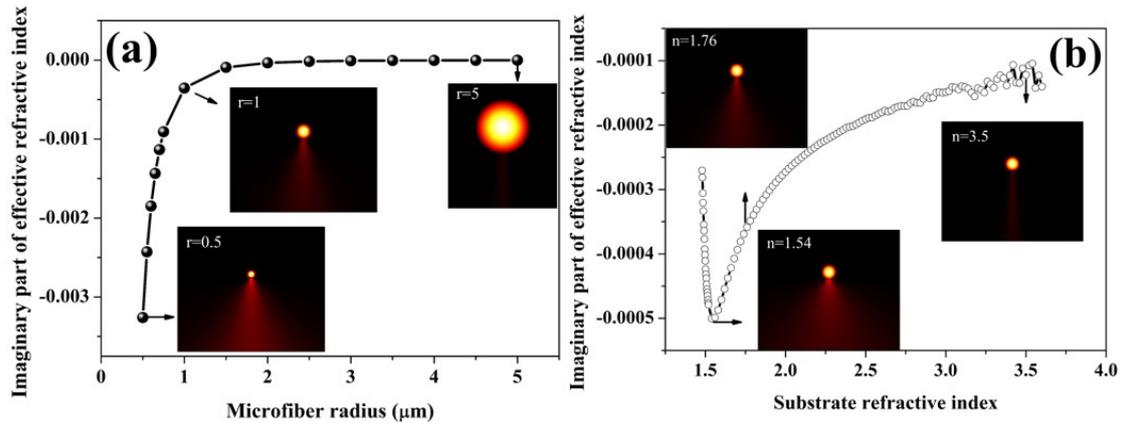

Fig.3

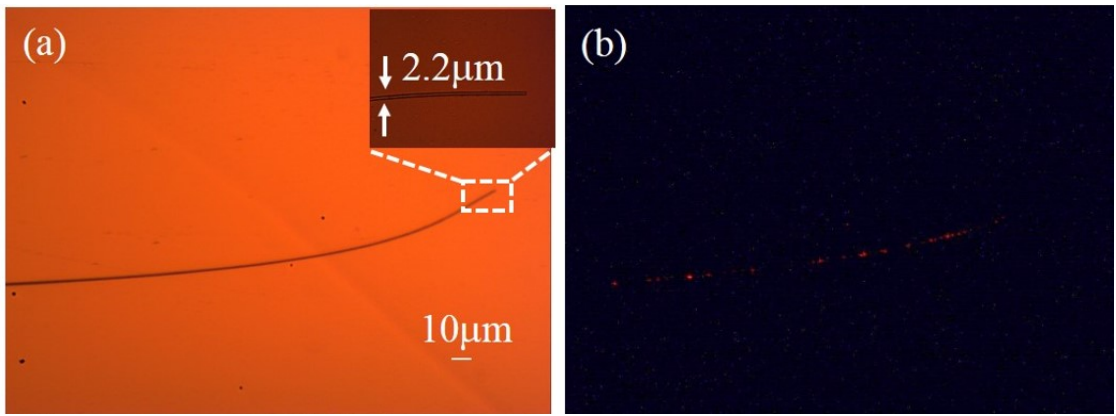

Fig.4

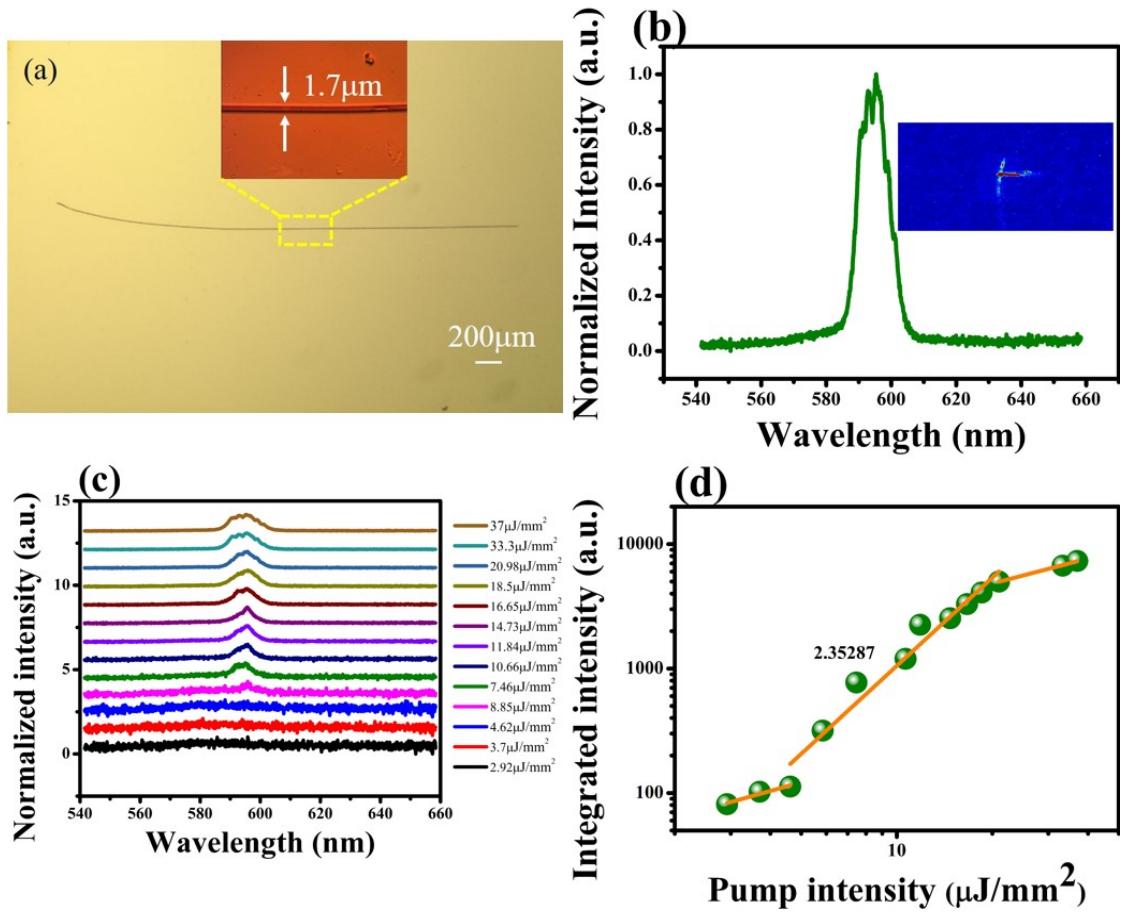

Fig.5

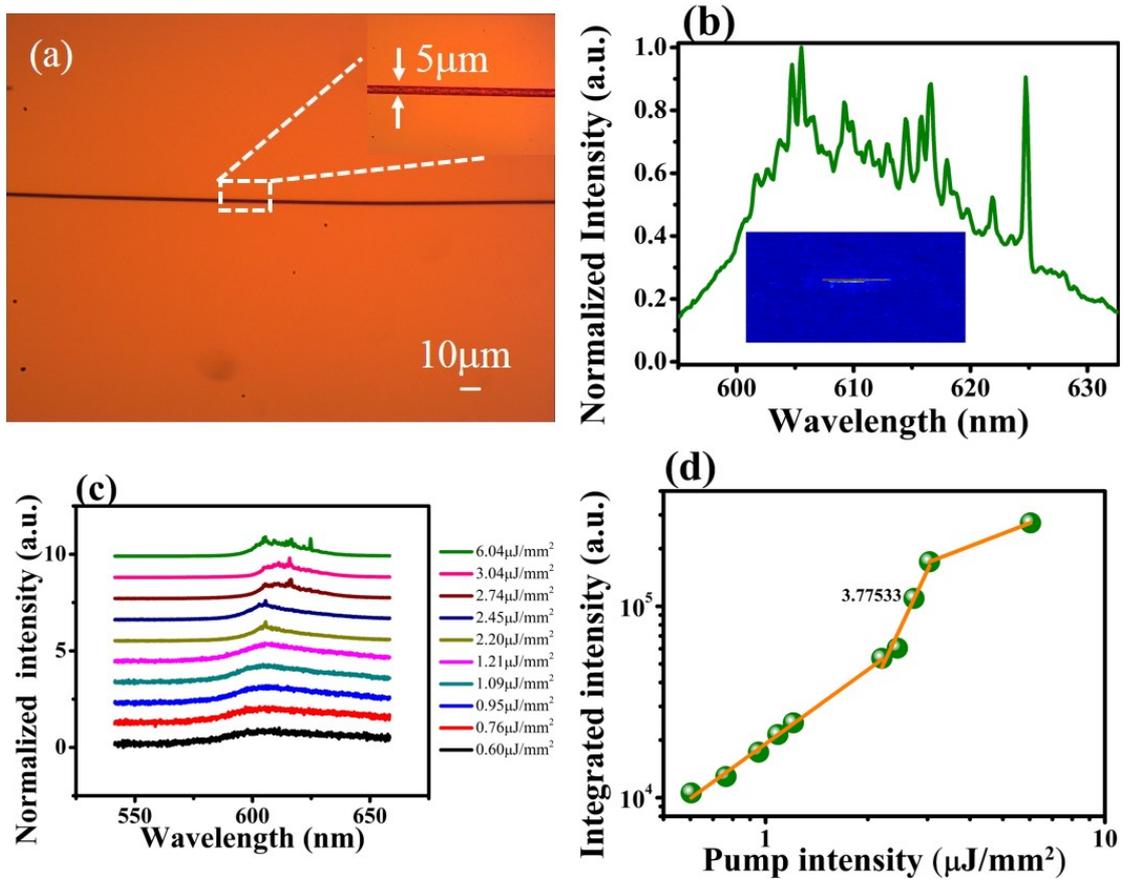

Fig.6